\documentclass[aps,pre,showkeys,showpacs,groupedaddress,amsmath,amssymb,floatfix,tw
ocolumn]{revtex4-1}
\usepackage{fmtcount} 
\usepackage{dcolumn}
\usepackage{bm}
\usepackage{epsfig}
\usepackage{amsfonts}
\usepackage{todonotes}
\usepackage{soul}
\usepackage{rotating}
\usepackage{color}
\usepackage{changes,hyperref}
\definechangesauthor[name={Deniz Eroglu}, color=blue]{DE}
\definechangesauthor[name={Norbert}, color=cyan]{NM}

\begin{document}

\title{Multiplex Recurrence Networks}
\author{Deniz Eroglu$^{1,2}$}
\email{eroglu@pik-potsdam.de}
\author{Norbert Marwan$^{1}$}
\author{Martina Stebich$^{3}$}
\author{J\"urgen Kurths$^{1,2,4}$}
\affiliation{
$^1$Potsdam Institute for Climate Impact Research (PIK), 14473 Potsdam, Germany\\
$^2$Department of Physics, Humboldt University, 12489 Berlin, Germany\\
$^3$Senckenberg Research Station of Quaternary Palaeontology Weimar, Am 
Jakobskirchhof 4, 99423 Weimar, Germany\\
$^4$Institute for Complex Systems and Mathematical Biology, University of Aberdeen, 
Aberdeen AB24 3UE, United Kingdom
}

\date{\today\\oo
Preprint version of \href{https://doi.org/10.1103/PhysRevE.97.012312}{doi:10.1103/PhysRevE.97.012312} }

\begin{abstract}
We have introduced a novel multiplex recurrence network (MRN) approach 
by combining recurrence networks with the multiplex network approach in order to 
investigate
multivariate time series. The potential use of this approach is demonstrated on coupled map
lattices and a typical example from palaeobotany research. In both examples, topological changes 
in the multiplex recurrence networks allow for the detection of regime changes in their
dynamics. The method goes beyond classical interpretation of pollen records 
by considering the vegetation as a whole and using the intrinsic similarity in the dynamics
of the different regional vegetation elements. We find that the different vegetation types behave 
more similar when one environmental factor acts as the dominant driving 
force.
\end{abstract}

\maketitle

\section{Introduction}
In order to understand the dynamical behavior of systems in a broad range
of scientific fields such as physics, biology, medicine, climatology, economy
etc., time series analysis provides crucial techniques. Although
investigation of time series can be done by various techniques, such as
basic statistics, symbolization, power spectra or similarity analysis, 
phase space based methods have become an
important role in dynamical systems' analysis. Recurrence of a
trajectory in its phase space is one of the most important fundamental
features of dynamical systems. Related approaches have been used for
several decades. The first recurrence based approach is known as the
Poincar\'e recurrence theorem introduced in 1890~\cite{poincare1890}.
The theorem indicates that almost all trajectories of dynamical
systems will turn back infinitesimally close to their previous positions
after a sufficiently long but finite time~\cite{poincare1890}.
Continuous dynamical systems can be defined by a set of ordinary differential
equations. These equation sets are called as flows and if the phase
space of a flow has a bounded volume, then the Poincar\'e recurrence
theorem is always valid~\cite{katok1995}. Among the many different
approaches of analyzing dynamical systems by their recurrence, the
recurrence plot (RP) is a multifaceted and powerful approach to study
different aspects of dynamical systems. Introduced by Eckmann et al.~in 1987,
the RP is a matrix to show the times of recurrences of a trajectory in its
phase space~\cite{eckmann1987}. Afterwards,  many statistical
quantification methods based on RP were developed for characterizing
dynamical properties, regime transitions, synchronization,
etc.~\cite{marwan2007}. 

Understanding underlying dynamics and detecting possible regime changes in the
evolution of dynamical systems are important problems studied by
time series analysis. For instance, we assume a dynamical system 
\begin{equation}
\vec{\dot x} =f(\vec{x},r)
\label{eq:ode}
\end{equation}
where $\vec{x}\in \mathbb{R}^m$, $f: \mathbb{R}^m \to  \mathbb{R}^m$ and a
control parameter ($r$) that has not to be constant on time $r=g(t)$. The purpose of
the analysis is to detect possible dynamical regime changes in the time series
caused by the time dependence of $r$. Transitions in the dynamics can
be detected by different RP based measures, which in general are
powerful to study complex, real-world
systems~\cite{trulla1996,marwan2002,donges2011}. Examples of their
successful application to real-world systems have been found in
medicine~\cite{jansen1991,kaplan1991,riley1999,marwan2002,neuman2009,carrubba2012}, Earth
science~\cite{richards1994,marwan2003,matcharashvili2008,donges2011,ozken2015,eroglu2016},
astrophysics~\cite{asghari2004,zolotova2009}, electrochemistry ~\cite{eroglu2014b},
and others~\cite{marwan2008,grassberger83,grassberger84}.

In the last decade, transformation of a time series to a complex
network has become a very powerful approach to analyze complex
dynamical systems. There are several ways to convert a time series to
a network such as symbolic dynamics based techniques~\cite{zhang2006,
small2013}, visibility graphs~\cite{lacasa2008,lacasa2009,lacasa2015}, cycle networks \cite{zhang2008netw}, or recurrence
networks (RN)~\cite{xu2008,marwan2009, donner2011}. In this work we consider
recurrence based approaches, since it is well known that 
recurrences are thumbprint of characteristic properties of dynamical
systems \cite{grassberger83, grassberger83c, afraimovich1997,
saussol2002, saussol2003a,donner2011epjb}. Moreover, recently we have presented a
work on fundamental behavior of recurrence plot measures which proves
that the Huberman-Rudnick unique scaling is valid for the RP measures,
while changing the control parameter of a given dynamical
system \cite{afsar2015}. The adjacency matrix of a complex network
represents the structure of the system and thus determines the links
between the nodes of a network. For unweighted and undirected
networks, the adjacency matrix is binary and symmetric, hence very
similar to an RP. As a consequence, we know that there is a similar
unique scaling behavior for RN measures and RNs have been used to
investigate real-world systems such as the climate
\citep{donges2011} or the cardio-respiratory system
\citep{ramirez2013b}.  

Naturally many real systems possess many degrees of
freedom and such systems can be described by multivariate time series.
Each component of these systems can be considered as a time series and
we can reconstruct a phase space with them. Meanwhile, when the number
of components are huge, we need to have longer time series for
enough occurrence of recurrences in the phase space in order to analyze system by the recurrences. However, in several
disciplines like astrophysics, earth sciences and economy, having long
time series cannot be ensured. Therefore the components are analyzed
one by one or some dimension reduction is applied, but these might result
in further information loss from the system. 

{An increasing number of dimensions requires longer time series for traditional recurrence networks. Usually, the period of a trajectory in the phase space is extending with an increasing number of dimensions. With other words, if the dimension of the phase space is increasing, we need longer and longer time series to have enough number of recurrences to analyze the given system. Therefore, traditional RNs are not efficiently applicable for short multivariate time series to interpret the behavior of the system, because of the scarcity of recurrences.} In order to gain the maximum yield from data sets which has a {many
degrees of freedom}, in this work  we introduce a multiplex network-based 
approach to analyze multivariate time series. Network modeling
is a main tool in characterizing a broad range of problems in physics,
biology, social interactions, Earth science, economy etc. If there are
interactions between components of the system they can be modeled by
a coupled network structure. For example, the RN approach has
been used to identify the direction of inter-systems relationships between bivariate time 
series \cite{feldhoff2012}. Now we represent each component of
the system as a separated RN and interpret it as a layer of a multiplex
network~\cite{bianconi2013, nicosia2013, dedomenico2013, kivela2014,
nicosia2015} that we call multiplex recurrence network (MRN). 
A similar approach based on visibility graphs was recently discussed by
Lacasa \textit{et al}.~\cite{lacasa2015}, resulting in multiplex visibility
graphs~\cite{lacasa2015}. Using the measures given by Lacasa
\textit{et al.}~\cite{lacasa2015}, we will show a comprehensive
comparison of regular RN and multiplex recurrence networks on 
coupled chaotic systems. By employing the recurrence properties
of the dynamical system for the multiplex network construction,
we focus on the dynamics, whereas the visibility graph is more 
restricted to the statistical properties of the system (e.g., Hurst exponent) 
\cite{donner2011,lacasa2009}.
We demonstrate our new method's efficiency
by investigating high dimensional systems which are not possible to be 
analyzed by the traditional RN approach directly. 

As a real-world application, we analyze a palaeoclimate record 
which is a multivariate data set of pollen taxa representing the variability of 
past vegetation in NE China over the Holocene and found significant variations
in the congruence of the vegetation dynamics of the considered tree species.

\section{Methods}

Recurrence based techniques have been successfully used for time
series analysis of physical, biological, economical, climate systems.
Multi-layer networks have been recently introduced as a powerful
representation of a specific network of networks. In this section, we
discuss RNs and how to use them in multi-layer networks in order to
analyze multivariate time series.

\begin{figure*}[htb]
\centering
\includegraphics*[width=1.0\linewidth]{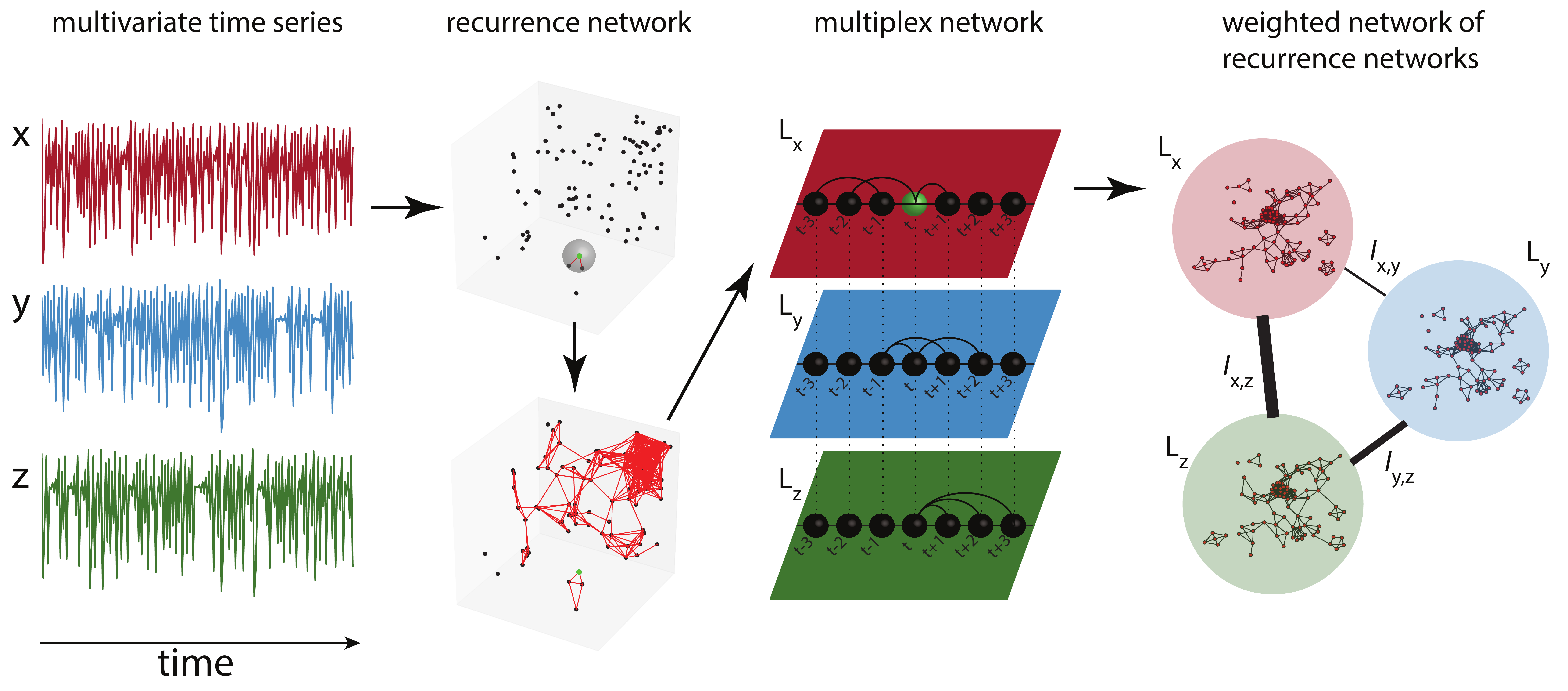}
\caption{Illustration of the procedure for generating network structures from a multivariate 
time series: \\ multivariate time series $\rightarrow$ recurrence networks $\rightarrow$ 
multiplex recurrence network $\rightarrow$ weighted network of recurrence networks. 
Dashed lines between multiplex network's layers connect all layers to all.}
\label{fig:multiplex_rn} 
\end{figure*}

\subsection{Recurrence Networks}

A time series $\{u_i\}_{i=1}^N$ can be reconstructed as a trajectory (Eq.~\ref{eq:ode})
in its phase space with time delay embedding \citep{packard80}
\begin{equation}
\mathbf{x}_i = (u_i, u_{i+\tau},...,u_{i+\tau (M-1)}),
\label{eq:embedding}
\end{equation}
where $M$ is the embedding dimension which can be found by a false
nearest neighbours approach and $\tau$ is the embedding delay which can be
computed by mutual information or auto-correlation \citep{kantz97}.

Two state vectors of the reconstructed time series are considered to
be recurrent if the second vector falls into the
neighbourhood (an $\epsilon$-radius sphere) of the first vector.
For the trajectory $\vec{x}_i$ $(i=1,
\ldots,N, \vec{x}_i \in \mathbb{R}^M)$, the adjacency matrix of RN,
$\mathbf{A}$, is defined as
\begin{equation}
A_{i,j}(\epsilon) = \Theta\bigl(\epsilon - \left\Vert \vec{x}_{i} - \vec{x}_{j} \right\Vert \bigr) -
\delta_{i,j},\;i,j=1,\ldots,N,
\label{eq:adjacency}
\end{equation}
where $N$ is the trajectory length, $\Theta(\cdot)$ is the Heaviside
function, $\left\Vert\, \cdot\, \right\Vert$ is the {Euclidian} norm and $\delta_{i,j}$ is the Kronecker 
delta ($\delta_{i,j}=1$ if $i=j$, otherwise $\delta_{i,j}=0$) \cite{marwan2009, eroglu2014}.

A RN is constructed in the following way: Consider
the time points of a time series as nodes of a network; if the nodes are
sufficiently close to each other, in other words, if the space vectors
are neighbours, then a link between them is drawn. $\mathbf{A}$ represents 
the network, where $A_{i,j}=1$ if $i$ and $j$ are connected,  otherwise $A_{i,j}=0$.

In all recurrence based applications, the threshold $\epsilon$ is an
arbitrarily selected small number. The selection of $\epsilon$ can
affect the results easily. In order to have a reasonable analysis,
some threshold selection techniques were proposed
\cite{ahlstrom2006b,marwan2007, eroglu2014}. However none of them is a certain way
to choose the threshold. To be consistent, in this work we always
used  the way which depends on the standard deviation of the time
series \cite{schinkel2008}.


\subsection{Multiplex and Weighted Recurrence Networks}

\textbf{Multiplex structure.} In this work, an $m$-layer multiplex
network is constructed by RNs. For $m$-dimensional
multivariate time series, we can create $m$ different RNs which have
the same number of nodes and each node is labeled by its associated
time. {These networks will form the different layers of a
multilayer network.} The layers are connected each other with the same time labeled
nodes. This procedure requires that the time sampling is the same for all of
the used time series. If a multilayer network consists of $m$-layers which has the same
number of nodes and the connections between layers are only between a node
and its counterpart in the other layers, then we call such networks
``multiplex". Networks, transformed from multivariate time series, are
compatible with the definition of multiplex networks, because each node is
uniquely assigned to a certain time point of the multivariate time series,
i.e., we will find the equally time-labelled nodes in all layers.

We consider an $m$-dimensional multivariate time series
$\{\bm{u}(t)\}_{t=1}^N$, with $\bm{u}(t) = (u_1(t), u_2(t), \ldots{},
u_m(t)) \in \mathbb{R}^m$ for any value of $t$. Then, the RN
of the $\kappa$th component of $\bm{u}(t)$ is created and located into the
associated layer~$\kappa$ of the multiplex network. For an example with 
$N=3$ time series these steps are illustrated in Fig.~\ref{fig:multiplex_rn}. First, we
have three time series and construct a phase space for each signal.
The RN of the $\kappa$th component of the time series is
calculated by Eqs.~(\ref{eq:embedding})--(\ref{eq:adjacency}) and then placed in to layer~$
\kappa$. We denote
the adjacency matrix of the $\kappa$th layer as
$A^{[\kappa]}={a_{ij}^{[\kappa]}}$ and $a_{ij}^{\kappa} = 1$ if nodes
$i$ and $j$ are connected in layer $\kappa$, $a_{ij}^{\kappa} = 0$ 
otherwise. The giant adjacency matrix describing the entire multiplex network is denoted by
\begin{equation}
\mathcal{A} =\left[\begin{array}{ccccc}
\mathbf{A}^{[1]} & \bf{I}$$_N & \hdots & \bf{I}$$_N \\ 
\bf{I}$$_N & \mathbf{A}^{[2]} & \ddots & \vdots \\
\vdots & \ddots & \ddots  & \bf{I}$$_N \\
\bf{I}$$_N & \hdots & \bf{I}$$_N & \mathbf{A}^{[m]}\end{array}\right]
\end{equation}
where $\bf{I}$$_N$ is the identity matrix of size $N$.

In order to measure the similarity between layer $\kappa$ and $\gamma$
of the MRN,  we use the {\it interlayer mutual information}
$I_{\kappa,\gamma}$ \cite{lacasa2015}:
\begin{equation}
I_{\kappa,\gamma} = \sum_{k^{[\kappa]}}\sum_{k^{[\gamma]}}P(k^{[\kappa]},k^{[\gamma]}) 
\log \frac{P(k^{[\kappa]},k^{[\gamma]})}{P(k^{[\kappa]})P(k^{[\gamma]})}
\label{eq:mutual_inf}
\end{equation}
where $P(k^{[\kappa]})$ and $P(k^{[\gamma]})$ are the degree
distributions of RNs at layer $\kappa$ and $\gamma$
respectively, and $P(k^{[\kappa]},k^{[\gamma]})$ is the joint
probability of the existence of nodes which has $k^{[\kappa]}$ degree at layer-$\kappa$
and $k^{[\gamma]}$ at layer-$\gamma$. The degree distribution $P(k)$ is a probability distribution function which holds a general structure information that how many nodes have each degree. The mutual information measures how much a system is similar to another. Instead of computing mutual information between original time series, degree distributions are considered in Eq.~\ref{eq:mutual_inf}. The difference to the commonly used direct mutual 
information of 
time series is that $I_{\kappa,\gamma}$ does not compare the probability of states 
(estimated from the time series) but 
the topological structure in the phase space based on the recurrences (using the recurrence 
matrix). This similarity measure $I_{\kappa,\gamma}$
quantifies the information flow between the multiplex networks and, thus, the
characteristical behaviour of the system. The average of the
quantity of $I_{\kappa,\gamma}$ over all possible pairs of layers of
MRN, gives a scalar variable $I=\left<I_{\kappa,\gamma} \right>
_{\kappa,\gamma}$ which captures the mean similarity of the degree
distributions of the RNs, i.e., the order of
coherence in the system. 


Another measure to quantify the coherence of the original multivariate 
system by the MRN is the {\it average edge overlap}:
\begin{equation} \omega =
\frac{\sum_i\sum_{j>i}\sum_{\kappa}a_{ij}^{[\kappa]}}{m\sum_i\sum_{j>i}\left(1-
\delta_{0,\sum_{\kappa}a_{ij}^{[\kappa]}}\right)}
\label{eq:omega} \end{equation} where $\delta_{ij}$ is the Kronecker
delta symbol. This measure {represents the 
average number of identical edges}
over all layers of the multiplex network~\cite{lacasa2015}. Like the 
interlayer mutual information Eq.~(\ref{eq:mutual_inf}), $\omega$ 
estimates the similarity and coherence with averaged existence
of overlapped links from node-$i$ to $j$ between all layers 
$\kappa$  and $\gamma$. {Note that $\omega$ can take values in the interval $[1/m,1]$, if the link between $i$ and $j$ occurs in only different layers $\omega=1/m$, i.e., $a_{ij}^{[\kappa]}=1$ and $a_{ij}^{[\gamma]}=0, \,\forall \gamma \ne \kappa$. If all links are identical in all layers then $\omega=1$. } 


\textbf{Weighted structure.} Another approach is the projection of all
layers onto one weighted network representation. Now we consider each
single layer of an MRN as a node and weighted edges between nodes
$\kappa$ and $\gamma$ are determined by the quantity $I_{\kappa,\gamma}$,  
Eq.~(\ref{eq:mutual_inf}). 

In order to quantify high-dimensional systems, converting multilayer systems to
weighted structures is computationally a very efficient approach.
The adjacency matrix of the multiplex network $\mathcal{A}$ is a
giant matrix ($Nm \times Nm$), but in the weighted network case the size
of the matrix is only $N\times N$. Furthermore, the quantification of
weighted networks is a well-developed
analysis~\cite{barrat2004,boccaletti2006}. Among many measures of
weighted networks, we use the {\textit{clustering coefficient}}
$\mathcal{C}_w$ and the \textit{average path length} $\mathcal{L}_w$ in
order to detect the transitions between different dynamical regimes.
The weighted network {clustering coefficient} is given by
\begin{equation}
\mathcal{C}_w = \frac{1}{m}\sum_{\kappa} \frac{1}{k_{\kappa}(k_{\kappa}-1)}\sum_{\gamma}
\sum_{\beta}(I_{\kappa,\gamma}I_{\kappa,\beta}I_{\gamma,\beta})^\frac{1}{3}
\label{eq:clustering}
\end{equation}
where $k_{\kappa}$ is the degree of node $\kappa$. The average
shortest path length of the weighted network is
\begin{equation}
\mathcal{L}_w = \sum_{\kappa,\gamma \in V} \frac{d(\kappa,\gamma)}{m(m-1)},
\label{eq:path_length}
\end{equation}
where $V$ is the set of nodes (layers of multiplex network) in the
weighted network and $d(\kappa,\gamma)$ is the weighted shortest path
length between nodes $\kappa$ and $\gamma$.

Both MRN measures, $I$ and $\omega$, represent the similarities in 
the linking structures of the RNs and have higher values for more similar ones.
For example, for periodic systems, even if there is phase difference, the corresponding RNs
would be rather similar, resulting in high values for $I$ and $\omega$. When the systems are 
more 
chaotic and have finally a more different recurrence structure, $I$ and $\omega$ will 
decrease.
This is similar for $\mathcal{C}_w$, since in periodic cases
the number of triangle structures in networks is increasing. However, it is opposite 
for $\mathcal{L}_w$ because the diameter of a denser network is in general smaller.


\section{Coupled Map Lattices}

As a first application, we consider multi-component dynamical systems,
namely coupled map lattices (CMLs), which are discrete-time models of
diffusively coupled oscillators on a ring model of $m$ sites. CMLs are
well-studied dynamical systems that model the behavior of non-linear
systems {and exhibit a variety} of phenomena~\cite{kaneko1992}. 

\begin{align}
x^{[\kappa]}_{t+1} = (1-\varepsilon)f(x^{[\kappa]}_t) + \frac{\varepsilon}{2} 
\bigl(f(x^{[\kappa-1]}_t) + f(x^{[\kappa+1]}_t)\bigr),
\label{eq:cml}
\end{align}

CMLs have been analyzed with various techniques to quantify their very
rich dynamics. Now we compare multiplex and regular RN
approaches for CMLs of $m=5$ diffusely coupled chaotic logistic maps
$f(x) = 4x(1-x)$, which show interesting dynamics in the range of the
control parameter $\varepsilon \in [0, 0.4]$ studied here
with an increment of $\Delta\varepsilon=0.005$. For instance, pattern
selections, high-dimensional chaos regimes, many different forms of
partially synchronized chaotic states occur. We compute a time series
of length $N =15,000$ for each value of $\varepsilon$. In order to
discard transients, we delete the first 10,000 values, resulting in
time series consisting of 5,000 values that have been used for all
analyses of the CMLs in this paper. All simulations of CMLs are
repeated and averaged over 100 realizations, since the system
strongly depends on the initial conditions. In order to
compare MRN and RN techniques, we construct a five-dimensional phase space
for regular RN and five separated single RN for each layer of
MRN.  

For consistency, we use a recurrence threshold that is
proportional to the standard deviation $\sigma$ of the data~\cite{marwan2007}.
In this work, we use $\epsilon=0.05\sigma$. {The logistic map is a 
one-dimensional dynamical system, therefore embedding is not required for applying 
recurrence networks in the layers of MRN.}

\begin{figure}[htb]
\centering
\includegraphics*[width=1.0\linewidth]{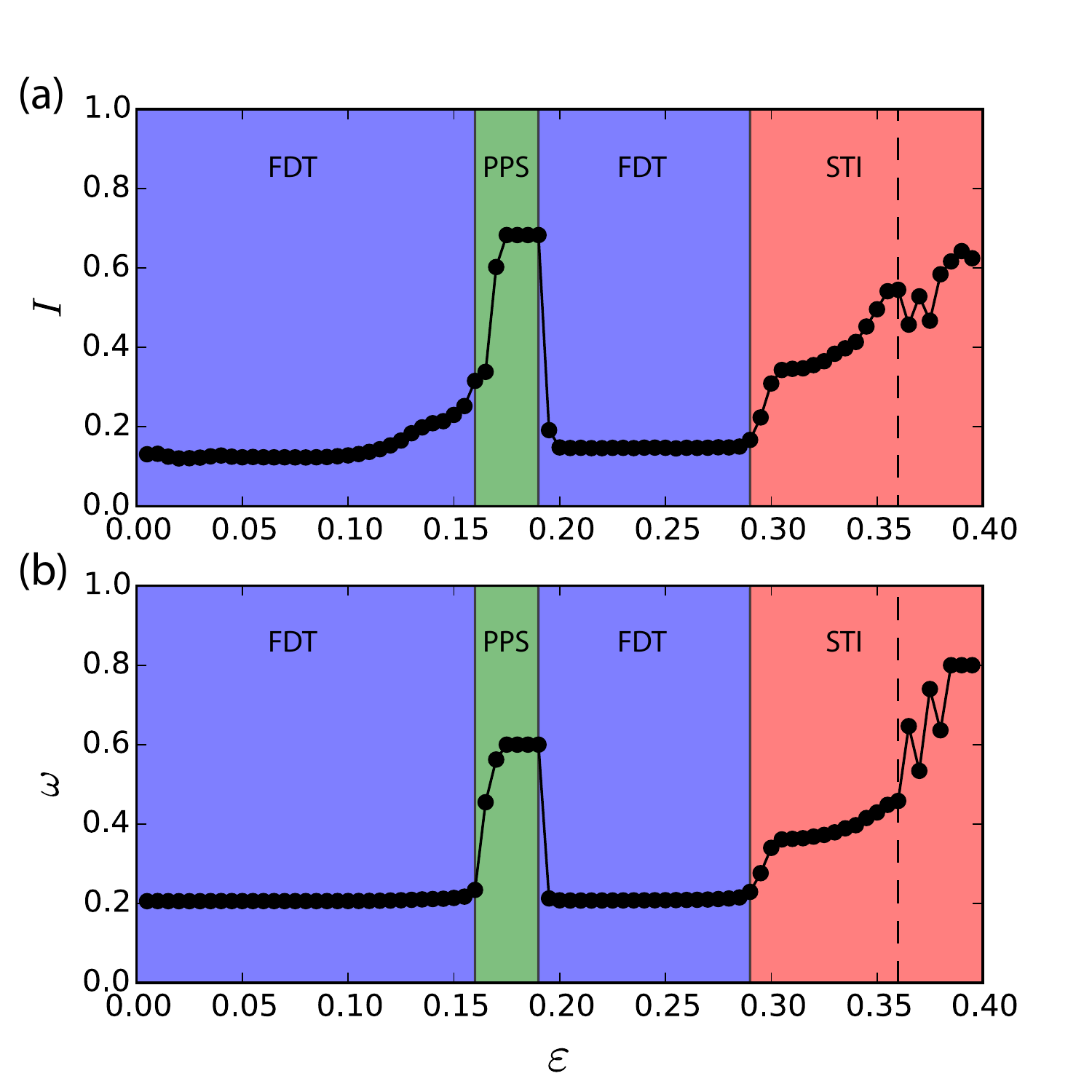}
\caption{Results for $m=5$ CMLs (Eq.~\ref{eq:cml}), (a) average mutual information ($I$), 
and (b) average edge overlap ($w$) of MRN.}
\label{fig:n5_res_multiplex} 
\end{figure}

\begin{figure}[htb]
\centering
\includegraphics*[width=1.0\linewidth]{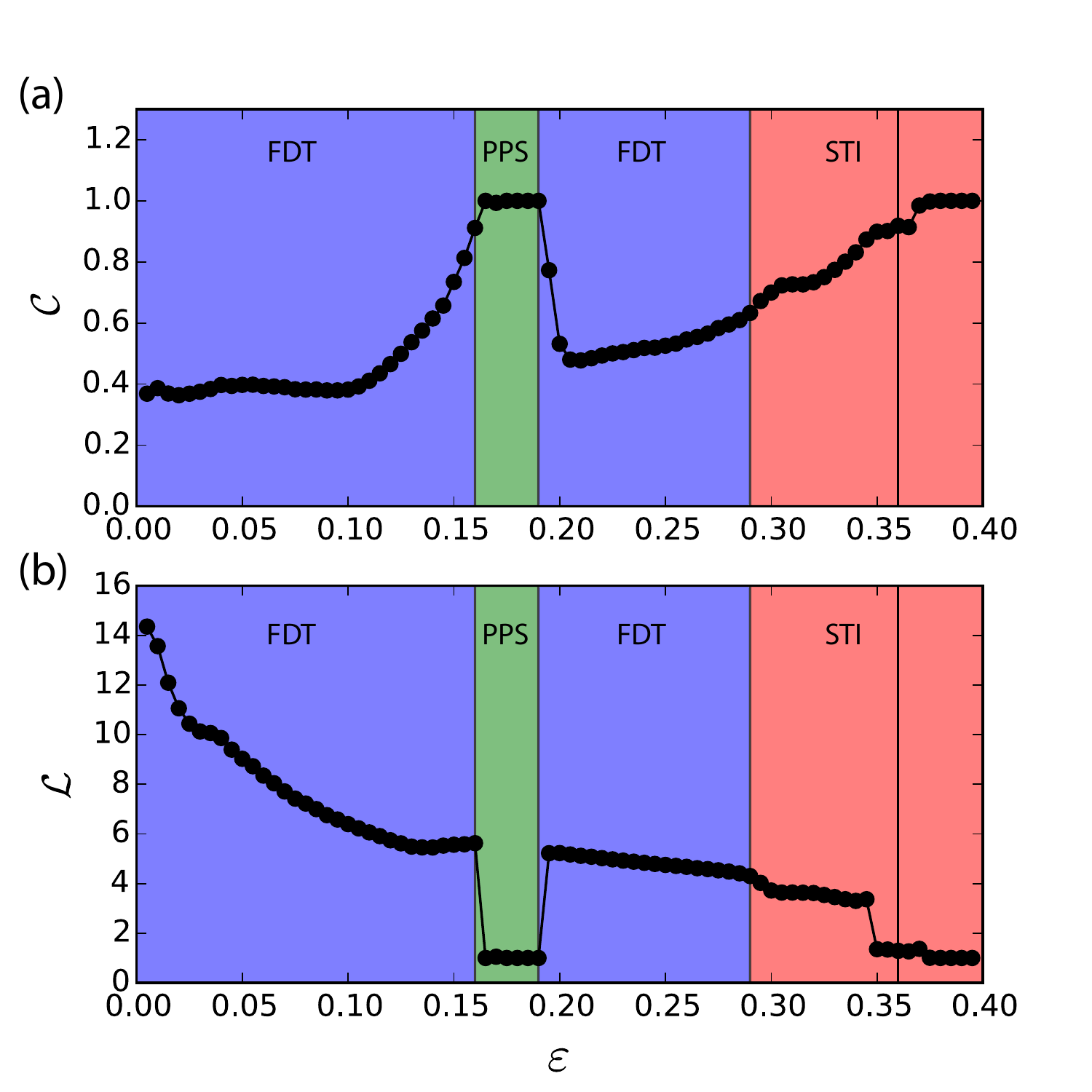}
\caption{Results for $m=5$ diffusively coupled map lattices CMLs, (a) {clustering coefficient} $
\mathcal{C}$, and (b) average path length $\mathcal{L}$ of \textcolor{blue}{the (regular)} recurrence network.}
\label{fig:n5_res_rn} 
\end{figure}

Although both techniques could detect the transitions from fully
developed turbulence (FDT),  $\varepsilon \in [0.0, 0.15]$ and
$\varepsilon \in [0.19, 0.285]$, to periodic pattern selection (PPS),
$\varepsilon \in [0.15, 0.19]$, the RN approach cannot distinguish the
transition from FDT to spatio-temporal intermittency (STI),
$\varepsilon \in [0.285, 0.4]$ (Figs.~\ref{fig:n5_res_multiplex},
\ref{fig:n5_res_rn}). Multiplex network's measures ($I$ and $\omega$)
can recognize every single transition and, especially $\omega$, can
distinguish splitting of trajectories into two attractors in STI at
$\varepsilon=0.36$ very clearly (Fig.~\ref{fig:n5_res_multiplex}). The
MRNs are more sensitive to regime changes than the RNs as well as the
MRNs are applicable on large system sizes when RNs are not suitable to
deal with them. 

\textbf{Large systems.} Simultaneous analysis of interacting
components of a system is very important for deep understanding of the
underlying dynamics. The RN approach is not an appropriate technique
to analyze such systems, because while the number of degrees of freedom
is increasing, the size (volume) of its phase space is getting larger
as well. Therefore, in order to analyze such systems, we need very
long data sets proportional to the size of the dimension of the system
for establishing recurrences. Although the recurrence based analyses
nicely deals with analyzing short time series, finding long data is a
common problem in time series analysis. However, MRNs overcome this
problem since they use each component of the system as a layer of the
giant network. 

\begin{figure}[htb]
\centering
\includegraphics*[width=1.0\linewidth]{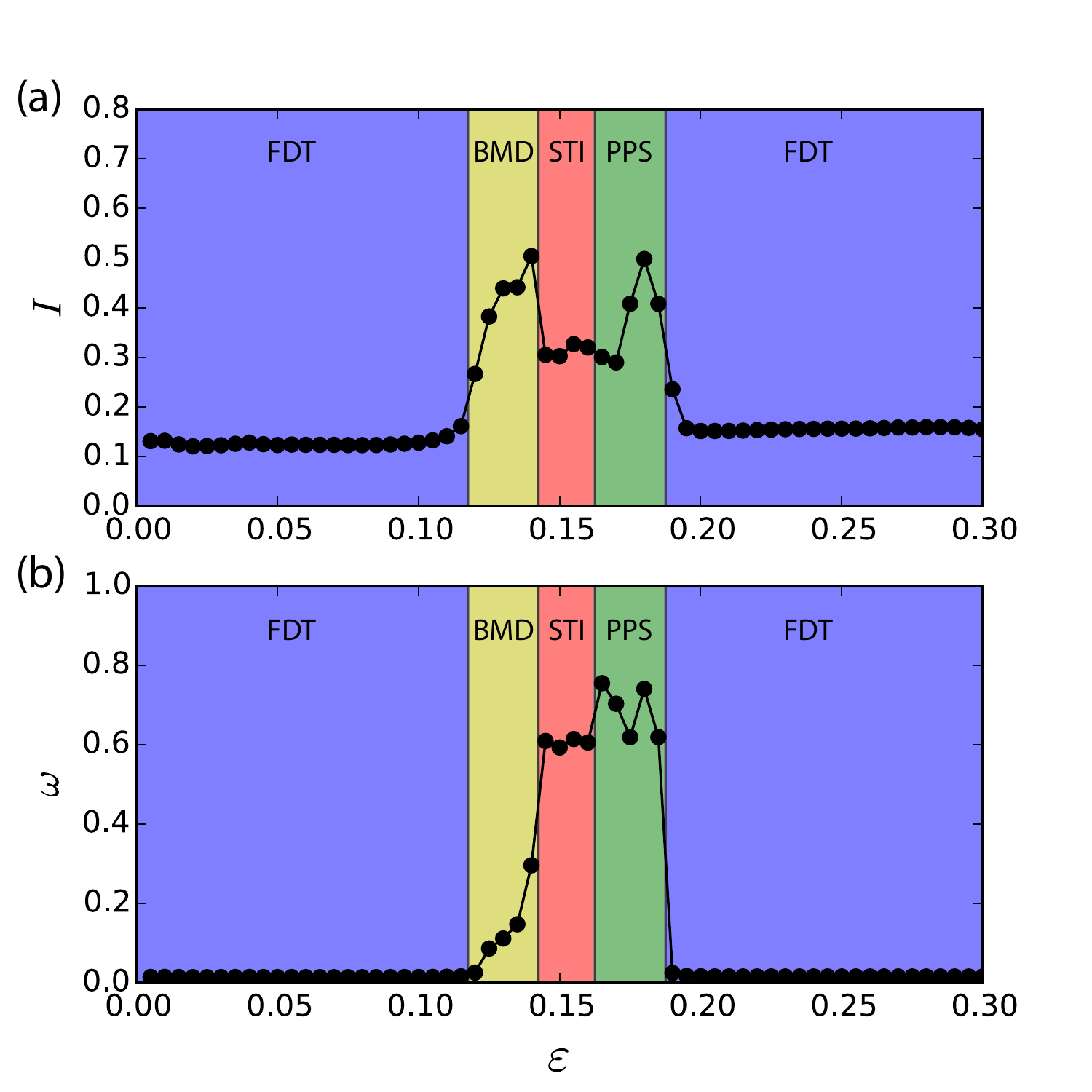}
\caption{Results for $m=200$ diffusively coupled map lattices CMLs (a) average mutual 
information ($I$), (b) average edge overlap ($w$) of multiplex network.}
\label{fig:n200_res_multiplex} 
\end{figure}

In this application, we use the same parameters of the RN and the
isolated dynamics as in the small system size one.
Fig.~\ref{fig:n200_res_multiplex} presents the results of the MRN for
$m=200$ diffusely coupled logistic maps, which possess one more
different regime than the small system, called Brownian motion of
defect (BMD), $\varepsilon \in [0.12,0.14]$. FDT occurs for
$\varepsilon \in [0.0,0.12]$ and $\varepsilon \in [0.19,0.30]$, STI is
observed for $\varepsilon \in [0.14,0.16]$ and PPS for $\varepsilon
\in [0.16,0.19]$. These regimes were observed and presented in detail
in \cite{kaneko1989,kaneko2001}. The MRN distinguishes the transitions between
different regimes clearly and this analysis is the first recurrence
based approach to detect these transitions in a large coupled system. 

\begin{figure}[htb]
\centering
\includegraphics*[width=1.0\linewidth]{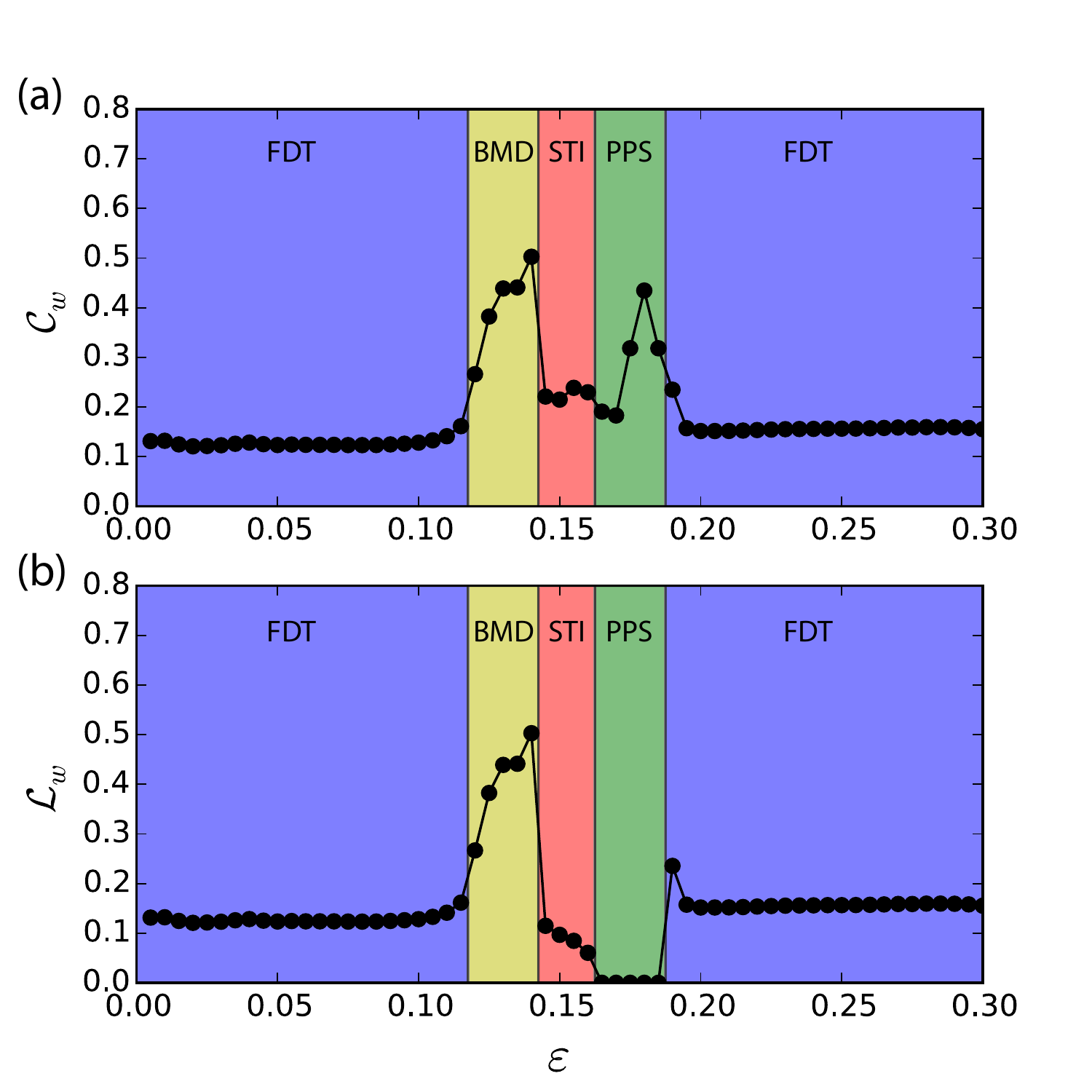}
\caption{Results for $m=200$ diffusively coupled map lattices CMLs (a) {clustering coefficient} 
of weighted network $\mathcal{C}_w$, (b) average path length of weighted network $
\mathcal{L}_w$}
\label{fig:n200_res_weighted} 
\end{figure}

Instead of analysing the giant adjacency matrix of MRN, we can
investigate the same system with taking advantage of the well-developed
weighted network analysis. Among the weighted network measures, the
{clustering coefficient} Eq.~(\ref{eq:clustering}) and the average
shortest path length Eq.~(\ref{eq:path_length}) are computed for the
associated network and the results are given in
Fig.~\ref{fig:n200_res_weighted}. As the results of MRN, the related
weighted network shows the transition very clearly. For large enough
multi-dimensional systems, both approaches can be used.    

The coupled maps are identical. In the fully developed chaotic
regime they are not synchronized, i.e., the evolution of the (chaotic) maps due 
to their distinct initial conditions is different. Varying the
coupling constant $\varepsilon$ changes the dynamical regime of the maps. According
to the high chaos in the FDT, each RN has a rather different topology compared to the
others, i.e., all the single layers of the MRN are quite different. The BMD is a less chaotic 
regime,
thus, the single layers of the MRN are less different than in the FDT regime (increasing the
inter-layer similarity). The PPS and STI regimes are closer to each other as both regimes 
exhibit periodic dynamics but at different scales.
Therefore, distinguishing the transition between PPS and STI is the hardest task.
Nevertheless, $I$ and $\omega$ are able to detect the corresponding regime transitions. 
However,
$\omega$ performs better, because it is a spatio-temporal measure, i.e., it
checks the existence of edges at the same locations in the different layers where the 
calculation for
$I$ does not take the nodes' identities into account.

\section{Real-world example -- multivariate palaeoclimate analysis}

So far we have been testing the performance of the new MRN method using 
prototypical models. As a real-world 
application with changing dynamics and with several variables, we treat a palaeoclimate 
problem. 
The investigation of the linkage between the climate conditions and specific environmental 
responses, 
as well as of changes in these relationships represent an important scientific challenge in 
palaeoclimatology and ecology in order to improve the understanding of climate impacts and 
feedback mechanisms. Information on past vegetation is useful to understand the 
palaeoecological 
processes linking climate and biosphere changes. 
Vegetation dynamics depends strongly on the environmental conditions and changes 
in the vegetation dynamics can, therefore, indicate critical changes in the environment 
(anthropogenic or climate impact).

\begin{figure}[htb]
\centering
\includegraphics*[width=\linewidth]{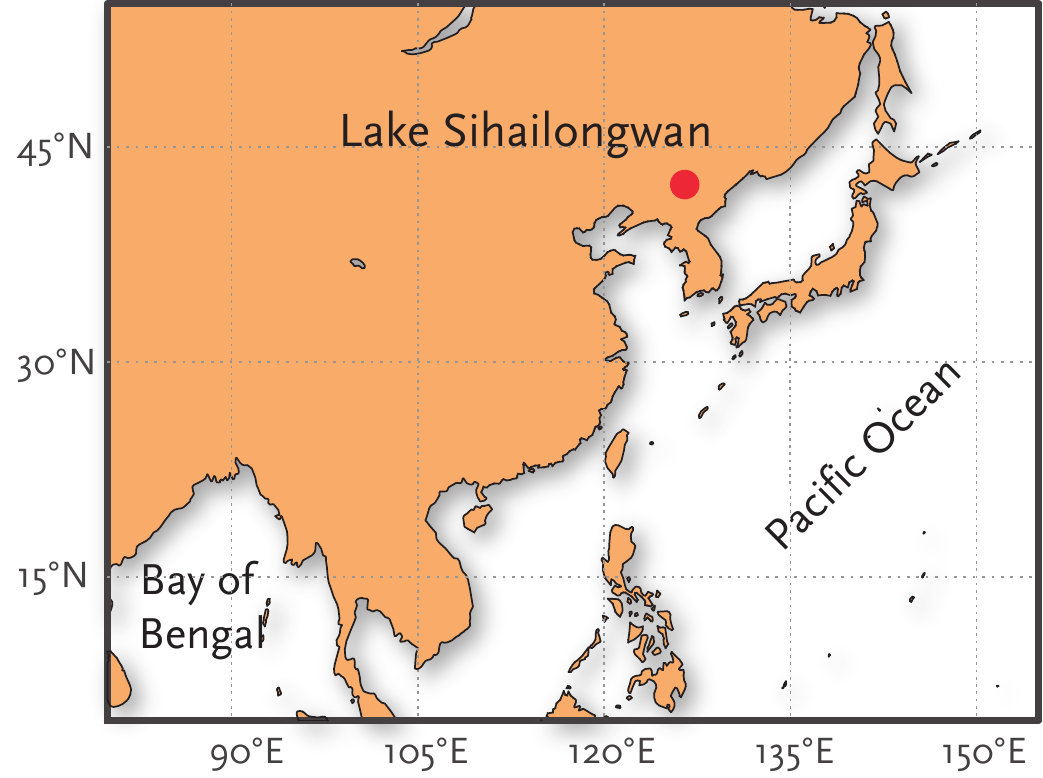}
\caption{Location of the Lake Sihailongwan from where the pollen
record for this study was sampled.}
\label{fig:map} 
\end{figure}

Pollen assemblages 
collected from lake sediments are commonly used 
proxies for palaeoecological and palaeoclimate investigations. 
Here we focus on the precisely dated Late Pleistocene-Holocene pollen record (last 16,500 
years (16.5~kyr BP)) 
from the Sihailongwan Lake  (42$^{\circ}$170$^\prime$ N,
126$^{\circ}$360$^\prime$ E), located in the Longgang volcanic field, 
Jilin Province, NE China (Fig.~\ref{fig:map}) \cite{stebich2009,stebich2015,stebich2015data}.

\begin{figure*}[htb]
\centering
\includegraphics[width=\linewidth]{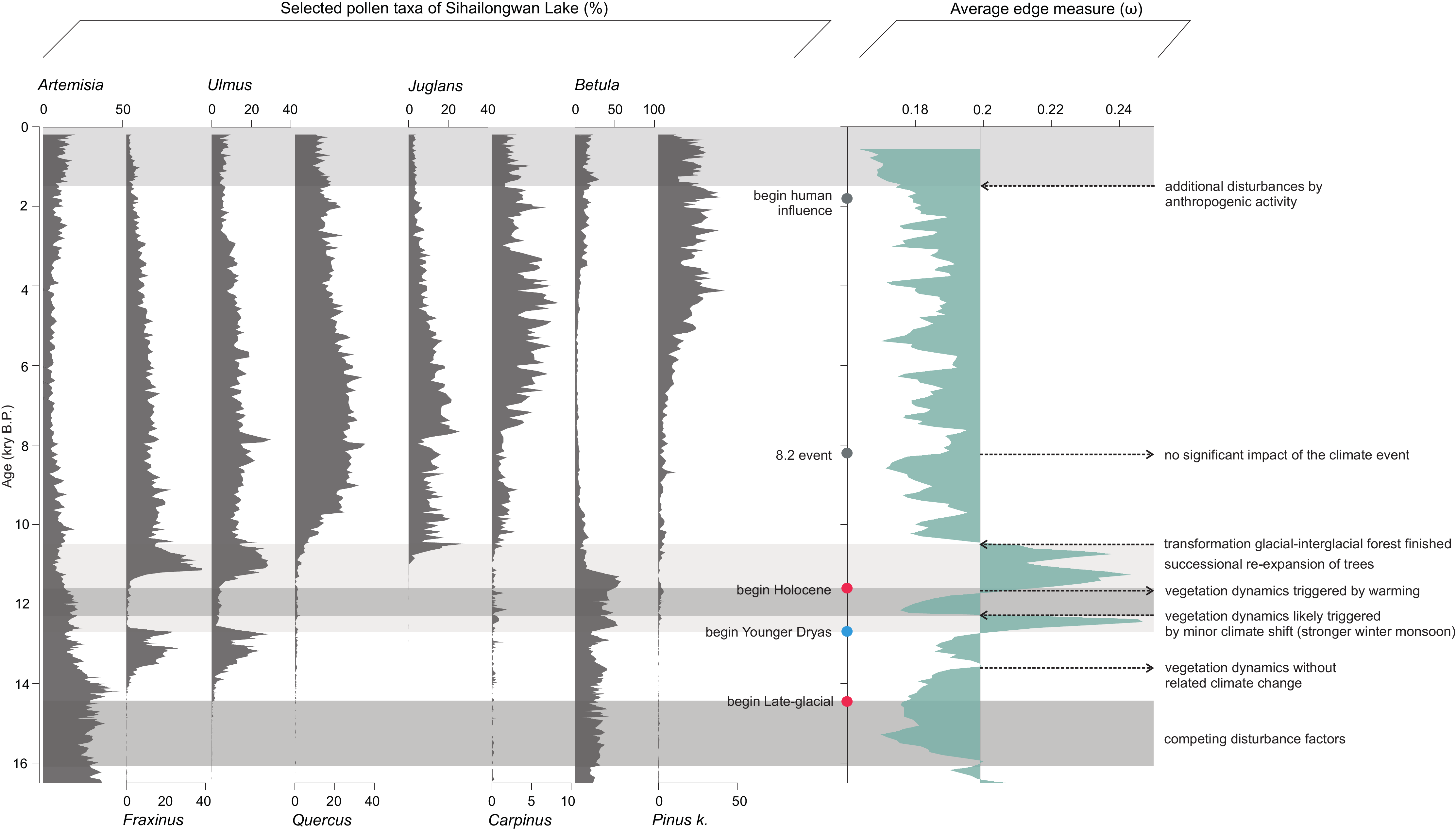}
\caption{
Pollen data set from Sihalongwan Lake (left) and derived average edge measure $\omega$ calculated in moving windows (right). 
The basevalue of the shading (0.199) marks the upper 95\% confidence interval, i.e., 
exceeding this value corresponds with highly
significant increase of $\omega$. The grey shaded bands mark
important environmental and climate transitions (see text).}
\label{fig:pollen_data} 
\end{figure*}

The Sihailongwan Lake is situated near the northern edge of the Asian summer monsoon 
system. 
Originally the Sihailongwan pollen record consists of 103 different pollen taxa, whereby 
several of them contain redundant information. 
We, therefore, selected seven tree pollen taxa ({\it Pinus koraiensis}, {\it Betula}, 
{\it Carpinus}, {\it Juglans}, {\it Quercus}, {\it Ulmus}, 
{\it Fraxinus} and one herbaceous genus ({\it Artemisia}), representing typical 
regional vegetation elements with different sensitivities on specific environmental conditions 
during the past 16.5 kyr BP
(Fig.~\ref{fig:pollen_data}). 

Since, the pollen record is irregularly sampled, we first 
interpolate the data into $N = 3,500$ points leading to the time resolution $\Delta t = t_{i+1} - 
t_i
\approx 4.95$ years $\forall\,i \in [0,N-1]$. In order to analyze the temporal
variation in the environmental dynamics, we apply a sliding window
approach consisting of 150 data points per window. By this choice, each
window covers about $740$ years and is suitable to
represent regime changes in the environmental dynamics. MRN is created for each
window one by one as the window slides over the time series with 90\%
overlap.

The application of the MRN technique to the multivariate pollen data set reveals
distinct fluctuations of $\omega$
(Fig.~\ref{fig:pollen_data}). A bootstrapping statistical test is applied to
evaluate these variations with respect to the null-hypothesis that multivariate
dynamics is constant over time within 95\% confidence \cite{marwan2013}. 

The considered measure $\omega$ quantifies the amount of similar dynamics in the
different pollen taxa. The higher $\omega$, the more similar vary the
different vegetation types in the sense of recurrence properties. This means,
it is not necessary that the pollen abundances go into the same directions, but
that, e.g., periodical variations are similar. Low $\omega$ values indicate, in
contrast, less similar behavior. This could happen, e.g., when the system is
disturbed and some vegetation populations recover faster than others. 

The pollen assemblages reveal relatively open, cold-dry adapted vegetation at Sihailongwan 
at the end of the pleniglacial. Starting at 16.0 kyr BP, we find a decrease of  $\omega$ until 
15.0 kyr 
BP. The apparent decrease of the similarity in the floristic response likely results from rather 
instable environmental conditions or competing factors influencing the vegetation dynamics. 
The instable environment at Sihailongwan is strongly evidenced by the sediment composition 
during this time interval. In particular, the occurrence of graded event layers with reworked 
soil material implies prevailing permafrost conditions with reduced seepage and erosion 
events during the final stage of the last glacial period \cite{stebich2009}. Moreover, the 
youngest Heinrich event (H1; 16.0 and 14.6 kyr BP), may have resulted in stronger winter 
monsoon \cite{wang2001}. By contrast, the ongoing change of CO$_2$ from the glacial to 
the 
interglacial level might have individually affected both  the physiology of the prevailing plants, 
and the vegetation dynamics in an increasingly less moisture-stressed environment 
\cite{Cowling1999,Monnin2004}.

Subsequent to the H1 event, the environmental conditions during the Late-glacial interstadial 
changed to warmer and moister climate, obviously related with a general change in the 
regional vegetation to one of more similar vegetation dynamics. The short positive excursion of 
$\omega$ at 13.6 kyr BP is, however, clearly not related to a climate shift, but most likely results from a 
change in interspecific competition after reaching a critical population density of {\it Fraxinus}
\cite{stebich2009}.

The Younger Dryas temperature decline about 12.7 kyr BP ago is one of the most prominent 
climate changes observed in Northern Hemisphere palaeoclimate records. Noticeable changes 
in the Sihailongwan pollen record provide a clear indication of a Younger-Dryas-like cooling in 
NE China beginning at 12.7 kyr BP. Unlike the abrupt termination of the Younger Dryas in the 
North Atlantic evidence, the more gradual changes of pollen assemblages indicate a 
successive substitution of the prevailing boreal woodland by species-rich broadleaf deciduous 
forests at the Sihailongwan lake during the first millennium of the Holocene (11.7 and 10.7 
kyr BP).

Our MRN analysis reveals significant high values of $\omega$ in the period between 12.7 and 10.5 kyr 
BP, which are interrupted by an abrupt drop in $\omega$ between about 12.3 and 11.7 kyr BP. The 
timing of this prominent change in vegetation dynamics corresponds to the recently detected 
bipartition of the Younger Dryas at Suigetsu Lake (Japan) \cite{schlolaut2017}, but neither the composition of 
pollen assemblages nor the sediments of the Sihailongwan sequence reveal a substantial 
environmental shift at 12.3 kyr BP. It is likely, that the ecosystem at Sihailongwan Lake may not 
be as sensitive to changes in winter conditions as the Japanese site \cite{schlolaut2017}, so that the climate 
shift may have influenced the intrinsic vegetation dynamics at Sihailongwan, but could not 
trigger substantial floristic and/or landcover changes. 

At approximately 10.7 kyr BP, the dynamic spread of thermophilous {\it Juglans} and a more gradual increase of 
{\it Quercus} mark a shift in the vegetation composition to oak- and walnut-rich mixed conifer hardwood forests. 
Subsequent to this transition, the MRN analysis data indicate persistent vegetation dynamics, but the general 
shift to lower $\omega$ values implies less similar behaviour of the pollen taxa. Interestingly, a shift in $\omega$ 
can be observed 
around 8.2 kyr BP, which might be related to the most prominent drop of the global temperatures during the 
Holocene. However, as its amplitude does not exceed that of other Holocene changes in vegetation dynamics, the 
shift does not prove a significant impact of the 8.2 event on the vegetation in NE China.

While substantial human impact cannot be traced by the pollen assemblages, monsoon changes, forest 
succession dynamics and flowering activity may therefore primarily drive the observed dynamics in the pollen 
values/vegetation at Sihailongwan throughout the Holocene. Nevertheless, weak indication of farming activity in 
conjunction with fires and modest grazing may explain the minor shift to lower Ï values from 1.8 yrs BP. 

Taken together, in our example we find that the patterns of vegetation dynamics are related to both intrinsic 
vegetation dynamics and external disturbance factors like climate changes, erosion or human impact. Obviously, 
the different vegetation types behave more similar when one environmental factor acts as the dominant driving 
force. This is particularly evident at the change to cooler and drier conditions at the beginning of Younger Dryas. 
Also the rapid mass expansion or the collapse of one or more tree species can have a similar effect on the 
dynamics of vegetation, as detected in our example at 13.6 kyrs BP and throughout the first millennium of the 
Holocene. On the other hand, less similar behaviour of the vegetation results from weaker or competing 
disturbance factors. 
Our example demonstrates that the new developed MRN technique goes beyond the classical 
interpretation of the pollen amplitude variation as a proxy of environmental conditions. 
A better understanding of this part of the climate-biosphere interaction is of crucial importance as non-linear 
feedback mechanisms and tipping points cause high uncertainty and an 
unpredictable future for humankind \cite{lenton2008,rockstrom2009}.


\section{Conclusions}

In this study we have introduced a novel multiplex recurrence network (MRN) approach 
which
allows us to analyze $m$-dimensional multivariate data simultaneously via joining
$m$ recurrence networks (RNs) together. Analyzing and understanding the characteristics of 
single
or low dimensional data with RNs is a known approach \cite{donges2011,donner2011epjb}. 
Nevertheless,
RN analysis of high dimensional systems is not trivial, because the increasing dimension of 
the phase space
would require longer time series. However, studies of real-world problems are often linked 
with short time series
and long time series are often not available. Our new approach considers each observation 
variable
in its own phase space and combines them later on. Therefore, it is
possible to analyze relatively short multivariate time series with MRNs.

Our extensive tests of the method have demonstrated that the MRN approach enables us to 
detect
abrupt regime transitions accurately in large dimensional systems. It can be
used to understand the underlying dynamical properties of prototypical models and
real-world applications. 

By analysing a pollen data set for the last 16 kyr, we are able to identify periods when
different vegetation types behave more similar and periods of less similar behaviour.
The changes in the vegetation dynamics coincide well with known climate transitions
and the increasing human impact. This analysis goes beyond the classical
interpretation of the pollen amplitude variation as a proxy of environmental conditions.
This new view can provide insights into plant communities and their dynamics
with respect to climate responses.

Especially for some research fields such as palaeoclimate
(pollen taxa, planktonic samples etc), neuroscience (short time crises in EEG),
economics (stock markets, currencies), where many dependent signals are
collected at the same time, our method provides a quantitative and objective way
to investigate their dynamics and detects hidden regime changes. 



\section*{Acknowlegments}
This work has been supported by German-Israeli Foundation for Scientific Research and 
Development 
(GIF), GIF Grant No.~I-1298-415.13/2015 and the European Union's Horizon 2020 Research 
and 
Innovation programme under the Marie Sk\l{}odowska-Curie grant agreement No.~691037 
(QUEST).

\bibliography{multiplex_rn}

\end{document}